\documentclass[final,3p,times]{elsarticle}

\usepackage{graphicx}
\usepackage{amsmath, amssymb}

\usepackage{amssymb}

\journal{Nuclear Physics B}

\newcommand{\beq}{\begin{equation}}
\newcommand{\eeq}{\end{equation}}
\newcommand{\bea}{\begin{eqnarray}}
\newcommand{\eea}{\end{eqnarray}}
\newcommand{\bev}{\begin{verbatim}}
\newcommand{\eev}{\end{verbatim}}
\newcommand{\nn}{\nonumber\\}

\newcommand{\BAN}{\begin{eqnarray*}}
\newcommand{\EAN}{\end{eqnarray*}}

\def\g5{\gamma_5}
\newcommand{\wideh}{}
\newcommand{\dov}{D_\mathrm{ov}}
\newcommand{\bomega}{{\boldsymbol \omega}}

\newcommand{\lp}{\left (}
\newcommand{\rp}{\right )}
\newcommand{\re}{\mathrm{Re}}
\newcommand{\im}{\mathrm{Im}}
\begin{document}

\begin{frontmatter}



\title
{New One-Flavor Hybrid Monte Carlo Simulation Method for Lattice Fermions with $\gamma_5$ Hermiticity.}

\author
{ Kenji Ogawa }

\address
{Institute of Physics, National Chiao-Tung University, Hsinchu 300, Taiwan}

\begin{abstract}
We propose a new method for Hybrid Monte Carlo (HMC) simulations with odd numbers of dynamical fermions on the lattice.
It employs a different approach from polynomial or rational HMC.
In this method, $\gamma_5$ hermiticity of the lattice Dirac operators is crucial and it can be applied to Wilson, domain-wall, and overlap fermions.
We compare HMC simulations with two degenerate flavors and $( 1 + 1 )$ degenerate flavors using optimal domain-wall fermions. 
The ratio of the efficiency, $ ( \mathrm{number~of~accepted~trajectories} ) / ( \mathrm{simulation~time} )$, is about 3:2. 
The relation between pseudofermion action of chirally symmetric lattice fermions in four-dimensional(overlap) and five-dimensional(domain-wall) representation are also analyzed.
\end{abstract}


\end{frontmatter}


\section{Introduction}
In hybrid Monte Carlo simulations\,\cite{Duane:1987de},
the positive-definiteness of the action is essential to consider it as the statistical weight.
When a lattice Dirac operator $D$ is given, 
a positive-definite action of two degenerate flavors is easily constructed by using the hermitian conjugate of $D$, namely $D^\dagger D$.
The major difference of two flavor simulations and $(2 + 1)$ flavor simulations is that 
one cannot easily write down the pseudofermion action for the one flavor sector.
Rational or polynomial HMC methods\,\cite{Borici:1995am,Clark:2006fx},
which approximates the square root of $D^\dagger D$, are mostly used for odd flavor simulations.

In this paper, we provide a pseudofermion action for 
the one flavor sector of the lattice fermions with $\gamma_5$ hermiticity without invoking the square root approximation for $D^\dagger D$.
%
The main idea is very simple. For any lattice Dirac operators $D$ with $\gamma_5$ hermiticity, $P_+ D P_+$ and $P_- (1/D) P_-$ are hermitian, and one can construct a one-flavor pseudofermion action using these.
The resultant action has the same determinant as $D$ without any approximations. 
The non-trivial parts are to first check the positive-definiteness of the pseudofermion action and the discussion of how to obtain the pseudofermion action when there are mass preconditioners like the one in Hasenbusch method.

The construction of the paper is as follows.
In Sec.\,\ref{sec:wilson} - \ref{sec:overlap},
the application to Wilson fermions, Wilson fermions with mass preconditioner, domain-wall type fermions and overlap fermions are demonstrated, respectively. 
In Sec.\,{\ref{sec:relation_4d5d}, the relation between the pseudofermion action in four and five dimensional representations are presented. 
Numerical results are given in Sec.\,\ref{sec:numerical}, while a
summary and conclusion are provided in Sec.\,\ref{sec:conclusion}.
\section{Wilson Fermions} \label{sec:wilson}
In this section, we derive a pseudofermion action which will yield the same
determinant as the Wilson-Dirac operator,
\beq
D_W(m)
= ( W + m ) {\bf 1} + \sum_\mu t_\mu \gamma_\mu
=
\begin{pmatrix}
( W + m ) { {\bf 1}_{2 \times 2} } & \sum_\mu t_\mu \sigma_\mu \\
\sum_\mu t_\mu \sigma_\mu^\dagger  & ( W + m ) { \bf 1 }_{2 \times 2} 
\end{pmatrix},
\label{eq:wilson}
\eeq
where
\beq
W = 
- \frac{1}{2} \sum_\mu 
\left [ U_{\mu}(x) \delta_{x + \hat \mu, y} + U^\dagger_{\mu}(x - \mu) \delta_{x - \hat \mu, y} \right ] + 4, 
\hspace{2mm}
t_\mu =
\frac{1}{2} 
\left [ U_{\mu}(x) \delta_{x + \hat \mu, y} - U_{\mu}^\dagger(x - \mu) \delta_{x - \hat \mu, y} \right ],
\nonumber
\eeq
\beq
\gamma_\mu = 
\begin{pmatrix}
0 & \sigma_\mu \\
\sigma_\mu^\dagger  & 0  
\end{pmatrix},
~~\sigma_\mu = ( i { \bf 1 }_{2 \times 2}, \sigma_i ),  
\nonumber
\eeq
${ {\bf 1}_{2 \times 2} }$ is two-by-two unit matrix
and $ \sigma_i \, (i = 1, 2, 3 ) $ are Pauli matrices.
Throughout this work, we use chiral basis for gamma matrices.
To get positive-definite pseudofermion action,
we use a determinant relation which is valid for general matrix, 
\beq
\det D_W(m) 
\cdot
\det \left ( P'_- {D_W(m)}^{-1} P'^\dagger_- \right )
=
\det \left ( P'_+ D_W(m) P'^\dagger_+  \right )
\eeq
where $P'_{+/-}$ are the projectors which reduce two chiral components into one chiral sector, $P'_+ = (1\,\, 0), P'_- = ( 0\,\,1) $.
The operators $(P'_- {D_W(m)}^{-1} P'^\dagger_- )$ and $ ( P'_+ D_W(m) P'^\dagger_+ )$ are hermitian due $\gamma_5$ symmetry.
When the inverse of $ (W+m) $ is well-defined, the determinant of (\ref{eq:wilson}) is written as
\bea
\det D_W(m)
=
\det \left ( P'_+ D_W(m) P'^\dagger_+  \right )
\cdot
\det \left ( \frac{1}{ P'_- {D_W(m)}^{-1} P'^\dagger_- } \right )
=
\det(W + m)^2 \det W_H(m),
\label{eq:DWH}
\eea
such that $ W_H(m) $ is the Schur complement of $ D_W(m) $, i.e.,  
\bea
W_H(m) = \frac{1}{P'_- {D_W}^{-1} P'^\dagger_-}
= ( W + m ) {\bf 1}_{2 \times 2} - \sum_{\mu, \nu} t_\mu \frac{1}{ W + m } t_\nu \sigma_\mu^\dagger \sigma_\nu .   
\eea
Thus, the pseudofermion action for one-flavor Wilson fermions can be written as 
\bea
S_{PF} 
= \Phi_1^\dagger (W + m)^{-2} \Phi_1 + \Phi_2^\dagger ( W_H(m))^{ -1 } \Phi_2  
= \Phi_1^\dagger (W + m)^{-2} \Phi_1 
- \Phi_2^\dagger P'_- H_W(m)^{-1} P'^\dagger_- \Phi_2
\label{eq:wls_spf}
\eea
where $H_W(m) = \gamma_5 D_W(m) $, $ \Phi_1 $ is a pseudofermion field without a Dirac index, 
and $ \Phi_2 $ is a pseudofermion field with two spinor components.

The positive-definiteness of the operator $W_H(m)$ are discussed as follows.
We note that  
for any background gauge field, the eigenvalues of $ W $ and $ (\sigma \cdot t) $ 
satisfy the inequalities\footnote{Throughout this work, $\lambda(X)$ means any one of the eigenvalues of an operator $X$.}:  ${ 0 \le \lambda(W) \le 8 }$, and 
${| \lambda(\sigma \cdot t) | \le 4 }$.
It then follows that $ W_H(m) $ is positive-definite for $ m > 4 $. 
Now, consider decreasing $m$ from 4 to a smaller value.
If the operator $W_H(m)$ is not positive for some $m$, e.g. $m'$,
then there must exist values of $m$ at which $ \det(W_H(m)) $ is zero or singular in the region $ m'< m < 4$.
Among these values, if we denote the largest one as $m_{cr}$,
then for $m > m_{cr}$, the positive-definiteness of $W_H(m)$ is assured.

The value $m_{cr}$ corresponds to the opposite sign of the smallest eigenvalue of $W$.
To see this, we use a relation\footnote{The proof of this relation is given in the appendix.},
\beq
\lambda_\mathrm{min}(W) \le \mathrm{Re} (\lambda(D_W(0))) \le \lambda_\mathrm{max}(W).
\label{eq:rlt_ev}
\eeq
From this relation, one can see that the value of $m$ 
which makes $\det \left ( W_H(m) \right ) $ singular (i.e. $ \det(W + m) = 0 $) is 
larger than the value of $m$ which satisfies $\det \left ( D_W(m) \right ) = \det \left ( W_H(m) \right ) = 0$.
There are no values of $m$ which satisfy $\det \left ( W_H(m) \right ) = 0$ or $\det \left ( W_H(m) \right ) = \pm \infty$ above that.
The condition for the positive-definiteness of $W_H(m)$ can then be written as
\beq
m_{cr} = m^{*}_0~~~{\rm s.t.}~~~ \det ( W + m^{*}_i ) = 0,~~ m^{*}_0 >  m^{*}_1 >  m^{*}_2 > \cdots, \nonumber
\eeq
\beq
\phi^\dagger \, W_H(m) \, \phi > 0 ~~{\rm~for~any~}\phi{\rm~and~for~} m > m_{cr}.
\label{eq:cnd_pd_wh}
\eeq
The smallest mass one can use in the one flavor method here is restricted by this bound. But it can be relaxed by using Hasenbusch's preconditioner as discussed in the next section.
%
%
%
\subsubsection*{Generating Pseudo Fermion Fields}
To generate the pseudofermion field $ \Phi_2 $ from a Gaussian random noise field $ \Xi_2 $,
we need to take the square root of $ W_H(m) $, i.e., $ \Phi_2 = \sqrt{ W_H(m) } \Xi_2 $ .
This can be approximated by using the rational function of $ W_H(m) $,
\beq
\Phi_2 = 
f_\mathrm{app} \left( W_H(m) \right )
\stackrel{\rm def}{=}
\left(p_0 + \sum_{l=1}^{N_\mathrm{app}} 
\frac{p_l}{( ( q_l + W + m ) {\bf 1}_{2 \times 2} - \sum_{\mu, \nu} t_\mu \frac{1}{W + m} t_\nu \sigma_\mu^\dagger \sigma_\nu } \right) \Xi_2 
\simeq
\sqrt{ W_H(m) } \, \Xi_2,
\label{eq:sqrtWH_Xi2}
\eeq
where $ p_0 $, $ p_l $ and $ q_l $ are expressed in terms of Jacobian elliptic functions. 
At first glance, the operations in (\ref{eq:sqrtWH_Xi2}) look formidable.
However, since $ W_H(m) $ is the Schur complement of $ D_W(m) $,
the inversion of the operator within the summation, in Eq.\,(\ref{eq:sqrtWH_Xi2}) 
can be obtained by the inversions of the operator $ (D_W(m) + q_l P_-) $, i.e.,  
\beq
\begin{pmatrix}
0 \\ \Phi_2 
\end{pmatrix}
=
P_- \left( p_0 + \sum_{l=1}^{N_\mathrm{app}} p_l \left ( D_W(m) + q_l P_- \right )^{-1} \right)
\begin{pmatrix}
0 \\ \Xi_2 
\end{pmatrix} .
\label{eq:sqrtWH}
\eeq
Note that one cannot apply the multi-shift solver in Eq.\,(\ref{eq:sqrtWH}), 
since $ P_- $ does not commute with $ D_W(m) $. 
However, for a given $N_\mathrm{app} $ number of inversions, 
the total number of iterations in the solver can be reduced
by using the same idea as the chronological inversion method \cite{Brower:1995vx}.
When one solves a set of linear equations $ ( D_W(m) + q_l P_+ ) \eta_l = \Xi_2 ~( 1 \le l \le N_\mathrm{app} ) $
with a given $\Xi_2$
serially from smaller $l$ by using the iterative method,
one can set a better initial guess $\eta^{(0)}_l$ for the iterative method to solve $ ( D_W(m) + P_+ ) \eta_l = \Xi_2 ~ ( 2 \le l ) $ 
by using a linear combination of the solutions $ \eta_j ~( j < l )$ which have already been calculated
i.e. $ \eta_l^{(0)} = \sum_{j < l} c_{lj} \eta_j $. The coefficients $ c_{lj} $ are determined according to the prescription written in Ref.\,\cite{Brower:1995vx}. 
 
In this one flavor method, generating pseudofermions using the approximation given by Eq.\,(\ref{eq:sqrtWH_Xi2}) makes the simulation not exact.
However, without using higher degrees of the approximation, 
one can make the algorithm exact by adding an accept/reject step after generating the pseudofermion field. 
In the exact algorithm, the pseudofermion field should be 
produced according to a probability distribution proportional to $ e^{- \phi^\dagger \frac{1}{W_H(m)} \phi} $.
This is obtained by multiplying the operator $\sqrt{W_H(m)}$ to a Gaussian noise field $\Xi$.  
In practice, however, the operator one uses in the simulation is approximated by $ f_\mathrm{app}(W_H(m)) $ rather than $ \sqrt{ W_H(m) } $.
This leads to a probability distribution that is proportional to $ e^{- \phi^\dagger \frac{1}{f_\mathrm{app}(W_H(m))^2} \phi} $.
To adjust the difference, one can add an accept/reject step for $ \phi $ with the probability
$ e^{- \phi^\dagger \left ( \frac{1}{W_H(m)} - \frac{1}{f_\mathrm{app}(W_H(m))^2} \right ) \phi} $.
This factor should be smaller than 1 and can be enforced
by choosing $ f_\mathrm{app}(x) > \sqrt{x} $ for the whole eigenvalue region of $W_H(m)$.
}
The discussion using eigenvectors is given in the appendix.

\section{Wilson fermions with the Hasenbusch method}\label{sec:wilson_hb}
The idea of the Hasenbusch method\,\cite{Hasenbusch:2001ne} is to factorize the determinant 
$ \det \left ( D_W ( m_1 ) \right )$ 
into a product of determinants,  
\beq
\det D_W( m_1 ) = \det \left ( D_W ( m_1 ) / D_W ( m_2 ) \right ) \cdot \det \left ( D_W( m_2 ) \right )
\eeq
and
the pseudofermion force coming from $\det\,(D_W(m_1)/D_W(m_2))$ is updated less frequently than the one coming from $\det\,(D_W(m_2))$ in the molecular dynamics steps.
The parameter $m_2$ is chosen such that the simulation cost is reduced.
Furthermore, as already mentioned, in the one flavor method for Wilson fermions presented here, the factorization of the determinant allows us to use a smaller fermion mass in the simulation. 
For the second factor $ \det( D_W(m_2) )$, the same argument as the previous section Sec.\,(\ref{sec:wilson}) can be applied. Here, we assume that $m_1 < m_2$ and that $m_2$ is large enough such that the positive-definiteness of $W_H(m_2)$ is always assured.  

Now, we consider how to treat the first factor $\det ( D_W( m_1 ) / D_W( m_2 ) )$.
Naively, one might consider the pseudofermion action like,
\beq
\phi^\dagger_1 \left ( \frac{ W + m_2 }{ W + m_1 } \right )^2 \phi_1
+
\phi^\dagger_2 \left ( W_H(m_2) \frac{1}{ W_H(m_1) } \right ) \phi_2
\eeq
but the second term is not hermitian in general, 
because $ W_H( m_1 ) W_H( m_2 ) \ne W_H( m_2 ) W_H( m_1 ) $ for $m_1 \ne m_2 $.
To remedy this situation,
we rewrite the determinant $\det(D_W(m_1))/(D_W(m_2)) $ by using an operator which has different masses for the chirality plus and chirality minus sector, i.e.,
\beq
\left [ \det \lp \frac{D_W(m_1)}{D_W(m_2)}\rp \right ]^{-1}
=
\det \left ( ( D_W(0) + P_+ m_1 + P_- m_2) \frac {1}{ D_W(m_1)} \right )
\cdot 
\det \left ( D_W(m_2) \frac {1}{D_W(0) + P_+ m_1 + P_- m_2 } \right ) .
\label{eq:det_hb}
\eeq
One may note that the first and the second factor on the right hand side are the same as the determinant of, 
\beq
1 + (m_2 - m_1) P'_- \frac{1}{D_W(m_1)} P'^\dagger_-,~~
1 + (m_2 - m_1) P'_+ \frac{1}{D_W(0) + m_1 P_+ + m_2 P_- } P'^\dagger_+,
\label{eq:opr_hb}
\eeq
respectively.
Then, the pseudofermion action is written as
\beq
S_{PF} 
= \Phi_3^\dagger \Phi_3 + \Phi_4^\dagger \Phi_4 
+ (m_2 - m_1) \Phi_3^\dagger P'_- \frac{1}{D_W(m_1)} P'^\dagger_- \Phi_3
+ (m_2 - m_1) \Phi_4^\dagger P'_+ \frac{1}{D_W(0) + m_1 P_+ + m_2 P_- } P'^\dagger_+ \Phi_4
\eeq
where $ \Phi_3 $ and $ \Phi_4 $ are pseudofermion fields with two spinor components.

The condition for the positive-definiteness is given as follows. When $m_1 = m_2$,
the operators in Eq\,(\ref{eq:opr_hb}) are the identity operators and their positive-definiteness is trivial.
When one decreases $m_1$ with $m_2$ fixed, the positive-definiteness will be lost
only if either of the determinants in Eq.\,(\ref{eq:det_hb}) becomes zero or singular. 
Thus, up to the largest $m_1$ which satisfies $\det(D_W(0) + m_1 P_+ + m_2 P_-) = 0$, the positive-definiteness of the operators in Eq\,(\ref{eq:opr_hb}) are assured.
Note that this mass value is smaller than the limit given in Eq\,(\ref{eq:cnd_pd_wh}).
This can be seen by using the eigenvalue relation which is similar to Eq\,(\ref{eq:rlt_ev}),
\beq
\lambda_\mathrm{min}(W+m_1) \le \mathrm{Re} (\lambda(D_W(0)+m_1 P_+ + m_2 P_-)) \le \lambda_\mathrm{max}(W+m_2).
\eeq

The largest $m_1$ which yields $\det( D_W(0) + m_1P_+ + m_2 P_- ) = 0$ is larger than the largest $m$ which yields $\det D_W(m) = 0 $. 
This can be understood from 
\beq
\gamma_5 ( D_W( 0 ) + m_1 P_+ + m_2 P_- ) = H_W \left (\frac{m_1 + m_2}{2} \right ) + \frac{m_1 - m_2}{2}
\eeq 
and the properties of spectral flows of $H_W(m)$\,\cite{Edwards:1998sh}.
Moreover, for some gauge configurations, $\det( D_W( m ) )$ may not be zero for any value of $m$. But there must exist $m_1$ which satisfies $\det( D_W(0) + m_1 P_+ + m_2 P_- ) = 0$ for any gauge configurations.
\section{Domain-wall fermions}\label{sec:domainwall}
The domain-wall type fermion operator\,\cite{Kaplan:1992bt} can be expressed as,
\beq
  \mathcal D_\mathrm{dwf}(m) = {\bomega} D_W( - m_0 ) (1 + c L(m)) + (1 - L(m))
  \label{eq:def_dwf}
\eeq
with $L(m)= P_+ L_+(m) + P_- L_-(m)$ such that,  
\beq
L_+(m)_{s,s'} =
\begin{cases} 
\delta_{s',s - 1},  &  1 < s \le N_s,   \\
- m \delta_{s', N_s}, &  s = 1, 
\end{cases}
\hspace{4mm}
L_-(m)_{s,s'} =
\begin{cases} 
\delta_{s',s + 1}, & 1 \le s < N_s,  \\
- m \delta_{s', 1},  & s = N_s,  
\end{cases}
\eeq
where $ m $ is the (bare) fermion mass, and $ m_0 \in (0,2) $ is a parameter called the "domain-wall height".
The label $s$ is the coordinate in the fifth dimension. Throughout this work, we assume that the number of sites in the fifth dimension $N_s$ is even.
The constant $c$ and the diagonal matrix $ \bomega = \mathrm{diag} \{ \omega_s \} $ specify the type of domain-wall fermion.
The operator $\mathcal D_\mathrm{dwf}(m) $ is the conventional domain-wall fermion for $c=0$ and $\omega_s = 1 $.
It becomes an optimal domain-wall fermion when $c=1$ and $\omega_s$'s  are tuned such that maximal chiral symmetry is obtained.

To obtain the pseudofermion action for one flavor,
we modify the operator $\mathcal{D}_\mathrm{dwf}(m) $ by multiplying $ 1/(1 + c L)$ from the right.
\beq
  \mathcal D'_\mathrm{dwf}(m) \stackrel{\rm def}{=} \mathcal{D}_\mathrm{dwf}(m) / ( 1 + c L ) = \bomega D_W( - m_0 ) + M(m,c)
\eeq
with
\beq
M( m , c ) = \frac{ 1 - L(m) }{ 1 + c L(m) } 
           = \frac{ 1 - L_+(m) }{ 1 + c L_+(m) } P_+ + \frac{ 1 - L_-(m) }{ 1 + c L_-(m) } P_-
           = M_+( m , c ) P_+ + M_-( m , c ) P_-.
\eeq
In the following, we suppress the argument $c$ of $M_\pm(m,c)$ for simplicity.
Using the Schur decomposition, the determinant of the operator $ \mathcal D'_\mathrm{dwf}(m)$ can be written as
\beq
 \det {\mathcal D'}_\mathrm{dwf}(m)
 =  
\det \left [ \bomega (W - m_0) + M_+(m) \right ]^2
\det {\mathcal W}_H(m)
 =
\det \left [ \bomega (W - m_0) + M_-(m) \right ]^2
\det \, \overline {\mathcal W}_H(m) , 
\eeq
where 
\beq
{\mathcal W}_H(m)
=
R_5 \left ( [ \bomega (W - m_0) +  M_-(m) ] {\bf 1}_{2 \times 2}
- t_\mu \frac{ 1 }{ \bomega ( W - m_0  ) +  M_+(m) } t_\nu \sigma_\mu^\dagger \sigma_\nu \right )
\eeq
\beq
\overline {\mathcal W}_H(m)
=
R_5 \left ( [ \bomega ( W - m_0 ) +  M_+(m) ] {\bf 1}_{2 \times 2}
- t_\mu \frac{ 1 }{ \bomega ( W - m_0 ) +  M_-(m) } t_\nu \sigma_\mu \sigma_\nu^\dagger \right ).
\eeq
Here $ R_5 $ is the reflection operator in the fifth dimension,  
$ \left( R_5 \right)_{s, s'}  = \delta_{s, N_s + s' - 1}$,  
which is introduced such that $ {\mathcal W}_H(m) $ and $ \overline {\mathcal W}_H(m) $ are hermitian.
For optimal domain-wall fermions, 
one still can choose $\omega_s$'s which maintain maximal chiral symmetry and satisfies 
$ \omega_{ N_s + 1 - s } = \omega_s $.
After incorporating the contributions of the Pauli-Villars fields, 
the fermion determinant for domain-wall fermions becomes,
\beq
\frac { \det {\mathcal D}_\mathrm{dwf}(m) }
{ \det {\mathcal D}_\mathrm{dwf}(1) }
 = 
\frac { 
\det \left[ \bomega ( W - m_0 ) +  M_+(m) \right]^2
\det {\mathcal W}_H(m) }{
\det \left[ \bomega ( W - m_0 ) +  M_+(1) \right]^2
\det \, \overline {\mathcal W}_H(1) 
}
\label{eq:dwf_fdet}
\eeq
In principle, one can use a pseudofermion action like, 
\beq
S_{PF}^{\rm (separate)} = 
\Phi_1^\dagger \left[ \bomega ( W - m_0 ) +  M_+(1) \right]\frac{1}{\left[ \bomega ( W - m_0 ) +  M_+(m) \right]^2} \left[ \bomega ( W - m_0 ) +  M_+(1) \right]\Phi_1 
+ \Phi_2^\dagger \frac{1}{{\mathcal W}_H(m)} \Phi_2 
+ \Phi_3^\dagger \overline {\mathcal W}_H(1) \Phi_3
\eeq
But it is known that in the two-flavor simulation of domain-wall fermions,
it is effective when a pseudofermion action uses a single set of pseudofermion field to estimate both the light fermion and Pauli-Villars terms\,\cite{Aoki:2004ht}.
Then, we use the same we used in Hasenbusch method for Wilson fermions.

Using the Schur decomposition of 
$ \left[ {\mathcal D}_\mathrm{dwf}(1) - \left (  M_+(1) -  M_+(m) \right ) P_+ \right] $,
we obtain the relation
\bea
\det \left [ \bomega ( W - m_0 ) +  M_+(m) \right ]^2
\cdot 
\det \left [ {\mathcal W}_H(m) +  \Delta_-(m) \right ] =
\det \left [ \bomega ( W - m_0 ) +  M_-(1) \right ]^2
\cdot 
\det \left [ \overline {\mathcal W}_H(1) -  \Delta_+(m) \right ], \nn 
\label{eq:det_lr1}
\eea
where
\bea
\label{eq:delta_m}
\left [  \Delta_+(m) \right ]_{s,s'} 
= \left [ R_5 \left (  M_+(1) -  M_+(m) \right ) \right ]_{s,s'},
\hspace{4mm} 
\left [  \Delta_-(m) \right ]_{s,s'} 
=\left [ R_5 \left ( M_-(1) -  M_-(m) \right ) \right ]_{s,s'}.
\eea
The properties of these matrices are given in the appendix.
Using (\ref{eq:det_lr1}), we can write the inverse of (\ref{eq:dwf_fdet}) as 
\beq
\det \left [
1 +
 \Delta_-(m)
\frac{1}{ {\mathcal W}_H(m) }
\right ]
\cdot
\det \left [
1 + 
 \Delta_+(m)
\frac{1}{
\overline {\mathcal W}_H(1) -  \Delta_+(m) }
\right ], 
\label{eq:det_dwf}
\eeq
and it can be used to construct the pseudofermion action.
Using (\ref{eq:rwt_delta}), we can simplify (\ref{eq:det_dwf}) to
\beq
\det \mathcal{A} \cdot \det \mathcal{B}
\label{eq:dwf_det}
\eeq
where, 
\beq
\mathcal{A} \stackrel{\rm def}{=} \left(1+ g'(m,1,c) \, (v^\dagger R_5 )_s \left[\frac{1}{{\mathcal W}_H(m)} \right]_{s,s'} ( R_5 v)_{s'} \right ) 
,~~
\mathcal{B} \stackrel{\rm def}{=} \left(1+ g'(m,1,c) \, (v^\dagger)_s \left[\frac{1}{\overline {\mathcal W}_H(1)- \Delta_+(m)} \right]_{s,s'} v_{s'} \right ).  
\eeq 
The constant $ g'$ and vector $ v $ are given in Eq.\,(\ref{eq:rwt_delta}) and (\ref{eq:def_v}).
In the following, the arguments of $g'$ are suppressed for simplicity ($ g' = g'(m,1,c) $).
Note that the five-dimensional matrix in Eq.\,(\ref{eq:det_dwf}) is reduced to a four-dimensional matrix in this expression.
Thus we can write the pseudofermion action for one-flavor domain-wall fermions as 
\bea
S_{PF} 
&=& \Phi_1^\dagger \, \mathcal{A} \, \Phi_1 + \Phi_2^\dagger \, \mathcal{B} \, \Phi_2 \nn
&=& \Phi_1^\dagger \Phi_1 - 
g'\,
\Phi_1^\dagger P'_-
(v^\dagger R_5 )_s \left [ \frac{1}{\gamma_5 R_5 {\mathcal D'}_\mathrm{dwf}(m) } \right]_{s , s'} (R_5 v)_{s'} 
P'^\dagger_-\Phi_1 
\nonumber \\
&+&  \Phi_2^\dagger \Phi_2
+
g'\,
\Phi_2^\dagger
P'_+
(v^\dagger)_s \left [ \frac{1}{ \gamma_5 R_5 {\mathcal D'}_\mathrm{dwf}(1) -  \Delta_+(m) P_+ } \right ]_{s , s'} v_{s'}
P'^\dagger_+
\Phi_2
\label{eq:pfaction_dwf}
\eea
where $ \Phi_1 $ and $ \Phi_2 $ are the pseudofermion fields (on the four-dimensional lattice) with two spinor components.
%
%

Now we assert that the operators in (\ref{eq:dwf_det}) are positive-definite for $ 0 < m \le 1 $. 
At $ m = 1 $, they are equal to the identity operator, and thus are positive-definite.
As $ m $ is decreased, 
the operators in (\ref{eq:dwf_det}) will cease to be positive-definite 
only if either of the determinants in (\ref{eq:dwf_det})
becomes zero or singular. 
Using $ [W, \frac{1}{\sqrt{\bomega}} M_-(1) \frac{1}{\sqrt{\bomega}}] = [W , \frac{1}{\sqrt{\bomega}} M_+(m) \frac{1}{\sqrt{\bomega}} ] = 0 $, and the fact that   
the eigenvalues of $\frac{1}{\sqrt{\bomega}} M_-(1) \frac{1}{\sqrt{\bomega}} $ and $\frac{1}{\sqrt{\bomega}} M_+(m) \frac{1}{\sqrt{\bomega}} $ have non-zero imaginary parts\footnote{ Recall that $M(m)$ for a conventional domain-wall fermion is a difference operator with anti-periodic boundary condition.}
for $ 0 < m \le 1$,
we immediately see that ${ ( \bomega ( W - m_0 ) +  M_-(1) ) }$ and ${ ( \bomega ( W - m_0 ) + M_+(m) ) }$
cannot have a zero eigenvalue for $ 0 < m \le 1 $.
Thus the operators in (\ref{eq:dwf_det}) are well-defined for $ 0 < m \le 1 $.
Furthermore, since (\ref{eq:det_dwf}) is equal to the determinant of the four-dimensional Dirac operator
with the approximation for the sign function of $ H_\mathrm{kernel} = \gamma_5 D_W ( 2 + (1 - c ) D_W)^{-1} $, the determinant cannot be zero and it follows that 
the operators in (\ref{eq:dwf_det}) are positive-definite for $ 0 < m \le 1 $.
\subsubsection*{Generating Pseudo Fermion Field}
We now discuss how to approximate the inverse square root of the operators 
$\wideh {\mathcal A}$ and $\wideh {\mathcal B}$ when one generates the pseudofermion fields $\Phi_1$ and $\Phi_2$ from the Gaussian noise.
Focusing on Eq.\,(\ref{eq:rlt_inv}),
we start from the inverse relation of five-dimensional operators,
\beq
\left [
\left ( \overline {\mathcal W}_H(m) + \alpha \wideh \Delta_+ \right ) 
/ 
\left ( \overline {\mathcal W}_H(m) + \beta \wideh \Delta_+ \right )
\right ]^{-1}
= 
\left ( \overline {\mathcal W}_H(m) + \beta \wideh \Delta_+ \right ) 
/ 
\left ( \overline {\mathcal W}_H(m) + \alpha \wideh \Delta_+ \right )
\eeq
\beq
\left (
1 + ( \beta - \alpha ) \wideh \Delta_+
\frac{1}{ \overline {\mathcal W}_H(m) + \beta \wideh \Delta_+ } 
\right )^{-1}
= 
1 + ( \alpha - \beta ) \wideh \Delta_+
\frac{1}{ \overline {\mathcal W}_H(m) + \alpha \wideh \Delta_+ }. 
\eeq
Multiplying by $ S^\dagger $ from the left and $ S $ from the right, one obtains
\bea
\left (
1 + ( \beta - \alpha ) g'
\mathrm{diag}(0, \cdots, 0, 1) 
{ \wideh S }^\dagger
\frac{1}{ \overline {\mathcal W}_H(m) + \beta \wideh \Delta_+ } 
\wideh S
\right )^{-1}
\nn= 
1 + ( \alpha - \beta ) g'
\mathrm{diag}(0, \cdots, 0, 1)  
{ \wideh S }^\dagger 
\frac{1}{ \overline {\mathcal W}_H(m) + \alpha \wideh \Delta_+ } 
\wideh S
\stackrel{\rm def}{=}\mathcal{M}.
\label{eq:relation_dwf}
\eea
Here, 
$ \mathcal{M}_{ss'} = 0  ~~{\rm for}~s < s' $.
Then, the relation, Eq.\,(\ref{eq:relation_dwf}), also holds for the sub-block $ (s,s') = (N_s, N_s) $,
\beq
\left (
1 + 
( \beta - \alpha ) \,
g'\,
{v}^\dagger  
\frac{1}{ \overline {\mathcal W}_H(m) + \beta \wideh \Delta_+ } 
v
\right )^{-1}
= 
1 + 
( \alpha - \beta ) \,
g'\,
{v}^\dagger  
\frac{1}{ \overline {\mathcal W}_H(m) + \alpha \wideh \Delta_+ } 
v.
\label{eq:rlt_inv}
\eeq
One can obtain a similar equation for ${\mathcal W}_H(m) $ and $\Delta_-$.

For the square root of a general positive-definite operator $A$, the rational approximation can be used.
\beq
\sqrt{A} \sim p_0 + \sum_{ i = 1 }^n \frac{ p_i }{ q_i + A }
\label{eq:rational}
\eeq
In the case of ${ \mathcal A} $, one has to calculate
\beq
\phi_1 = 
\left (
1 + 
g'\,
{v}^\dagger R_5
\left [
\frac{1}{ {\mathcal W}_H(m) }
\right ]
R_5 v
\right )^{ - \frac{1}{2} } ~\Xi_1
= 
\left (
1 - 
g'\,
{v}^\dagger R_5
\left [
\frac{1}{ {\mathcal W}_H(m) + \Delta_- }
\right ]
R_5 v
\right )^{ \frac{1}{2} } ~\Xi_1.
\eeq
Each term of the summation in Eq.\,(\ref{eq:rational}) is 
\bea
\frac{ p_i }
{ q_i + 1 - g'\, v^\dagger R_5 \frac{ 1 }{ {\mathcal W}_H(m) + \Delta_- } R_5 v } 
& = &
\frac{ p_i }{ 1 + q_i } 
\left [
 1 + \frac{1}{ 1 + q_i } 
g'\,
v^\dagger R_5 
\frac{ 1 }{ {\mathcal W}_H(m) + \frac{ q_i }{ 1 + q_i } \Delta_- } R_5 v 
\right ]
\\
& = & 
\frac{ p_i }{ 1 + q_i } 
+
\frac{ p_i }{ ( 1 + q_i )^2 } 
g'\,
v^\dagger R_5 
\frac{ 1 }{ {\mathcal W}_H(m) + \frac{ q_i }{ 1 + q_i } \Delta_- }
R_5 v 
\\
& = & 
\frac{ p_i }{ 1 + q_i } 
+
\hat p_i
\,
g'\,
v^\dagger R_5
\frac{ 1 }{ {\mathcal W}_H(m) + \hat q_i \Delta_- }
R_5 v 
\eea
Then, to obtain $\phi_1$ one needs to calculate 
\beq
\sqrt{ \frac{1}{ { \mathcal A} } } \, \Xi_1 \sim \hat p_0 \, \Xi_1
 + \sum_i^n \hat p_i \, 
g'\,
v^\dagger R_5
\frac{ 1 }{ {\mathcal W}_H(m) + \hat q_i \Delta_- } 
R_5 v \, \Xi_1 
\eeq
with,
\beq
\hat p_0 = p_0 + \sum_i^n \frac{ p_i }{ 1 + q_i } ,
~~~
\hat p_i = \frac{ p_i }{ ( 1 + q_i )^2 } ,
~~~
\hat q_i = \frac{ q_i }{ 1 + q_i }. 
\eeq
The same method can be used for the operator $\mathcal{B}$.
\section{Overlap Fermions}\label{sec:overlap}
In this section, 
we construct the pseudofermion action for one flavor overlap fermions\,\cite{Neuberger:1997fp},
\beq
D_\mathrm{ov}(m) = 1 + m + \left ( 1 - m \right ) \gamma_5 \mathrm{sign}( H_W (-m_0)). 
\eeq
This operator satisfies Ginsparg-Wilson Relation(GWR)\,\cite{Luscher:1998pqa},
\beq
\dov(m)^\dagger \dov(m) = (1-m^2) \left ( \dov(0)^\dagger + \dov(0) \right ) +  4 m^2 
= 2(1-m^2) \left( P_+ \dov(0) P_+ + P_- \dov(0) P_- \right ) + 4 m^2.
\eeq
The overlap fermion also possesses $ \gamma_5 $ hermiticity, 
and so the application is similar to Wilson fermions.
We begin by breaking this operator into its chiral components.
\beq
D_\mathrm{ov}(m) 
=
\begin{pmatrix}
P'_+ D_\mathrm{ov}(m) P'^\dagger_+
&
P'_+ D_\mathrm{ov}(m) P'^\dagger_-
\\
P'_- D_\mathrm{ov}(m) P'^\dagger_+
&
P'_- D_\mathrm{ov}(m) P'^\dagger_-
\end{pmatrix}
\stackrel{\rm def}{=} 
\begin{pmatrix}
D_\mathrm{++}(m) & D_\mathrm{+-}(m) \\
D_\mathrm{-+}(m) & D_\mathrm{--}(m) 
\end{pmatrix}
\eeq

The determinant is written as,
\beq
\det \left ( D_\mathrm{ov}(m) \right )
= 
\det \left ( D_{++}(m) \right ) 
\det \left ( D_{--}(m) - D_{-+}(m) \frac{1}{D_{++}(m)} D_{+-}(m) \right ) 
\label{eq:det_dov}
\eeq

By using GWR, one can show that the operators on the right hand side are positive-definite for $ 0 < m \le 1$, provided that $D_\mathrm{ov}(m)$ itself is well defined.
The pseudofermion action is written as
\beq
\phi_1^\dagger \frac{1}{D_{++}(m)} \phi_1
+ 
\phi_2^\dagger
P'_-
\frac{1}{D_\mathrm{ov}(m)} 
P'^\dagger_-
\phi_2
\eeq
Practically, 
one has to use 
a reflection/refraction\,\cite{Fodor:2003bh} or topology fixing term\,\cite{Fukaya:2006vs}
to treat or avoid singularities in $ D_\mathrm{ov}(m) $
related to the topological change.

The authors of \cite{DeGrand:2006ws} proposed a simulation method for one-flavor  
with overlap fermions. 
We now highlight the differences between their work and our work.
In Ref.\,\cite{DeGrand:2006ws}, the authors use GWR, 
and factorize $ \det {\dov(m)}^\dagger \dov(m)$ into two parts.
\beq
\det {\dov(m)}^\dagger \dov(m) = \mathrm{det} P'_+ \dov(m)^\dagger \dov(m) {P'_+}^\dagger \cdot \mathrm{det} P'_- \dov(m)^\dagger \dov(m) {P'_-}^\dagger 
\eeq
The difference of the first factor and second factor of the right-hand side 
comes from the topological zero-mode of $\dov(m)$.
Then, the one-flavor determinant is written as
\beq
\mathrm{det} \dov(m) 
= \mathrm{det} P'_+ \dov(m)^\dagger \dov(m) {P'_+}^\dagger \cdot m^{(N_+ - N_-)} 
= \mathrm{det} P'_- \dov(m)^\dagger \dov(m) {P'_-}^\dagger \cdot m^{(N_- - N_+)}.
\label{eq:fact_dov_1}
\eeq 
Here, $N_{+/-}$ are the numbers of topological zero-modes with a definite chirality. 
By using GWR, one obtains the relation between these operators and Schur complement,
\beq
P'_- \dov(m)^\dagger \dov(m) {P'_-}^\dagger =  
D_{--}(m) \left ( D_{--}(m) - D_{-+}(m) \frac{1}{D_{++}(m)} D_{+-}(m) \right ).
\eeq 
In other words, in Eq.\,(\ref{eq:det_dov}), the determinant is factorized as,
\beq
\det \left ( \dov(m) \right ) = 
\det \left ( D_{++}(m) \right ) 
\cdot 
\det \left ( \frac{P'_- \dov(m)^\dagger \dov(m) {P'_-}^\dagger }{D_{--}(m)} \right ) 
=
\det \left ( D_{++}(m) \right ) 
\cdot 
\det \left ( \frac{ 2( 1 + m ) D_{--}(m)  - 4 m }{D_{--}(m)} \right ) 
\label{eq:fact_dov_2}
\eeq
Later, we show that the second factor corresponds to the $\mathcal A$ term in Eq.\,(\ref{eq:pfaction_dwf}).
Due to the cancellation between the numerator and the denominator,  
the force for the $\mathcal A$ term in HMC is smaller than that of the $\mathcal B$ term.

When one performs a HMC simulation with a topological fixing term, 
it is apparent that using the pseudofermion action with the factorization in Eq.\,(\ref{eq:fact_dov_1})
is more effective than the factorization in Eq.\,(\ref{eq:fact_dov_2}).  
For the five-dimensional representation Eq.\,(\ref{eq:pfaction_dwf}), using only the $\mathcal{B}$ term
and choosing the fermion mass parameter $m'$ to satisfy $ ( 1 - m' ) = ( 1 - m^2 ) / m $,
one can perform the HMC simulation respecting the factorization in Eq.\,(\ref{eq:fact_dov_1}). 
\section{The relation of the pseudofermion action with four and five dimensional representations}\label{sec:relation_4d5d}
The ratio of determinants, Eq.\,(\ref{eq:dwf_fdet}), is equivalent to 
the determinant of the effective four-dimensional operator,
\beq
D_{4d}(m) = \frac{1}{2} ( 1 + m ) + \frac{1}{2}( 1 - m )\gamma_5 f(H_\mathrm{kernel}(-m_0)),
\eeq
where the function $ f(x) $ is polar or a rational function which approximates the sign function. The form of the function $f$ is determined by $\bomega$. 
The operator $ H_\mathrm{kernel} $ is defined as,
\beq
H_\mathrm{kernel}(-m_0) = \gamma_5 \frac{ D_W(-m_0) }{2 - (1 - c) D_W(-m_0)}.
\eeq 

By using the Schur decomposition, the one flavor pseudofermion action is written as
\beq
\Phi_1^\dagger P'_- \frac{1}{ D_{4d}(m) } P'^\dagger_- \Phi_1  + \Phi_2^\dagger \frac{1}{ P'_+ D_{4d}(m) P'^\dagger_+} \Phi_2
\label{eq:pfaction_4d}.
\eeq
Examining two cases, $c = 0$ and $c = 1$, we show that the five-dimensional expression Eq.\,(\ref{eq:pfaction_dwf}) reproduces
Eq.\,(\ref{eq:pfaction_4d}). 
This equivalence is not only of theoretical value, 
but aids in practical simulations.
When one generates $\phi_1$ and $\phi_2$,
from Eq.\,(\ref{eq:pfaction_dwf}),
one has to know the eigenvalue spectrum of $\mathcal A$ and $\mathcal B$, a priori,
in order to apply the rational approximation for the square root function. 
However, in the four-dimensional case, the spectrum is already known.
%
\subsubsection*{case 1: $\bf c = 0$}
In this case, the propagator of the five-dimensional fermion at the boundary $s = 1 {\rm \,or \,} N_s$ yields 
the propagator of the four-dimension fermion\,\cite{Chiu:2003bv},
\beq
D_{ch} + m \stackrel{\rm def}{=}
\frac{ (1 - m) D_{4d} }{ 1 - D_{4d} } 
= \frac{ 1 + \gamma_5 f( H_\mathrm{kernel}(- m_0) ) }{1 - \gamma_5 f(H_\mathrm{kernel}(- m_0)) } + m
\label{eq:def_sinv}
\eeq
\beq
\frac{1}{D_{ch} + m} = B \frac{1}{\mathcal{D}_\mathrm{dwf}(m)} R_5 B^\dagger.
\label{eq:def_s}
\eeq
Here, $B$ is defined as
\beq
B_{s} = P_- \delta_{1,s} + P_+ \delta_{N_s,s} 
\eeq
This relation holds only for $c=0$. Note that this operator satisfies chiral symmetry $D_{ch} \, \gamma_5 + \gamma_5 \, D_{ch} = 0$ in the limit $N_s \rightarrow \infty$, in which $f(x)$ becomes the sign function.
From Eq.\,(\ref{eq:def_sinv}) and Eq.\,(\ref{eq:def_s}),
\beq
P_- \left ( \frac{1}{D_{4d}(m)} \right ) P_-
 = P_- + (1-m) P_- \left [ \frac{1}{\mathcal{D}_\mathrm{dwf}(m)} \right ]_{s=1,s'=N_s} P_-. 
\eeq
After incorporating the pseudofermion field,
the left-hand side becomes the first term of Eq.\,(\ref{eq:pfaction_4d}).
The right-hand side equals to the $\mathcal A$ term of Eq.\,(\ref{eq:pfaction_dwf})
by substituting $g' = (1-m)$ and $ v^\dagger = ( 0, \cdots , 0 , 1 )$.

Next, consider the $\mathcal B$ term. 
By using Eq.\,(\ref{eq:rlt_inv}), one obtains,
\bea
& & 1 + (1-m) v^\dagger \frac{1}{\overline {\mathcal W}_H(1) - \Delta_+(m) } v 
=
\left [ 
1 - (1-m) v^\dagger \frac{1}{\overline {\mathcal W}_H(1) } v
\right ]^{-1}
\nn
&=& 
\left [ 1 - (1-m) P'_+ \frac{1}{D_{ch} + 1} P'^\dagger_+ \right ]^{-1}
= \left [ P'_+ D_{4d} P'^\dagger_+ \right ]^{-1}
\eea
where we used the relation $D_{4d}(m) = (D_{ch} + m)/(D_{ch} + 1) $.
\subsubsection*{case 2: $\bf c = 1$}
In this case, there are relations for $D_{4d}$ and five-dimensional operators\,\cite{Chiu:2002ir},
\beq
\left [ B \, {\mathcal D}_\mathrm{dwf}(m)^{-1} \, {\mathcal D}_\mathrm{dwf}(1) \, B^{\dagger} \right ] = D_{4d}(m)^{-1} 
\label{eq:relation_l1},
\eeq
\beq
\left [ B \, {\mathcal D}_\mathrm{dwf}(1)^{-1} \, {\mathcal D}_\mathrm{dwf}(m) \, B^{\dagger} \right ] = D_{4d}(m) 
\label{eq:relation_l2}.
\eeq

By using the relations ,
\beq
B ( 1 + c L(m) )^{-1} = \frac{ \sqrt{\lambda_c } }{1 + c^4 m} \, ( P_+ v^\dagger + P_- v^\dagger R_5 ),
\eeq
\beq
\Delta(m) \, ( 1 + c L(m) ) B^\dagger = g(m,1,c) \sqrt{\lambda_c }\, ( P_+ v + P_- R_5 v ),
\eeq
one can show that Eq.\,(\ref{eq:relation_l1}), Eq.\,(\ref{eq:relation_l2}) and projector $P_{+/-}$ reproduce the operators in Eq.\,(\ref{eq:pfaction_dwf}).
\section{Numerical tests}\label{sec:numerical}
We compare the efficiency of the HMC simulation for two-flavor and $ (1+1) $-flavor QCD with 
domain-wall type fermion with $c = 1$ and $\omega_s=1$, 
on a $ 12^3 \times 24 \times 16 (N_s) $ lattice.  
For the gluon action, we use Iwasaki gauge action at $ \beta = 2.30 $.
In the molecular dynamics, we use the Omelyan integrator \cite{Takaishi:2005tz}, 
and the Sexton-Weingarten method \cite{Sexton:1992nu}.
The pseudofermion action for the two-flavor simulation is the one with even-odd preconditioning 
which is described in Ref.\,\cite{Chiu:2009wh}.
The time step for the gauge field, ($\Delta \tau_\mathrm{Gauge}$), is the same 
for both two-flavor and $(1+1)$-flavor cases, while 
the time step ($\Delta \tau_\mathrm{PF}$) for the pseudofermion fields 
in the $(1 + 1)$-flavor case is four times larger than that for the two-flavor case.
The acceptance rate is roughly the same for both cases. 
We use conjugate gradient (CG) with mixed precision for the inversion of 
the quark matrix (with even-odd preconditioning). The length of each trajectory is set to two.
After discarding 300 trajectories for thermalization, we accumulate 100 trajectories for 
the comparison of efficiency. Our results are given in Table.\ref{table:test1}.
We see that the acceptance rate is almost the same for $(1+1)$-flavor and two-flavor simulations. 
If the auto-correlation time is the same, then the efficiency of HMC can be estimated by 
the total acceptance divided by the CG iteration number, 
and the efficiency ratio for two-flavor and $(1+1)$-flavor is about $ 3:2$. 

\begin{table}
\begin{center}
\begin{tabular}{|c|c|c|c|c|c|c|} \hline
$ m $ &
$N_\mathrm{f}$ &
$N_\mathrm{Iter}^\mathrm{(HB)} / 10^{3} $ &
$N_\mathrm{Iter}^\mathrm{(MD)} / 10^{3} $ &
$N_\mathrm{Iter}^\mathrm{(Total)} / 10^{3} $ & Acceptance &
$ \mathrm{Acceptance}/N_\mathrm{Iter}^\mathrm{(Total)}$  \\ \hline
$0.019$ & $1 + 1 $ &  75\,(1) & 260\,(1) & 345\,(1) & 0.88\,(3) & $ 2.6\,(1) \times 10^{-6} $ \\ \cline{2-7}
        & $ 2      $ & 0.6\,(1) & 239\,(2) & 240\,(2) & 0.90\,(3) & $ 3.8\,(2) \times 10^{-6} $ \\ \hline
$0.038$ & $1 + 1 $ &  54\,(1) & 125\,(1) & 179\,(1) & 0.90\,(3) & $ 5.0\,(2) \times 10^{-6} $ \\ \cline{2-7}
        & $2       $ & 0.6\,(1) & 112\,(1) & 113\,(1) & 0.91\,(3) & $ 8.0\,(3) \times 10^{-6} $ \\ \hline
\end{tabular}
\caption{Comparison of HMC efficiency for the 2-flavor and $(1+1)$-flavor QCD with optimal domain-wall quarks. 
The step size for the gauge field 
$\Delta \tau_\mathrm{Gauge}$  
is $ 0.007 (0.010) $ for $ m = 0.019 (0.038) $. 
while the step size 
$\Delta \tau_\mathrm{PF}$ 
for $ (1+1) $-flavor pseudofermions is 
$ 0.14 (0.20) $ for $ m = 0.019 (0.038) $, 
which is 4 times larger than that for the 2-flavor case. 
Here, $N_\mathrm{Iter}^\mathrm{(HB)}$, $N_\mathrm{Iter}^\mathrm{(MD)}$, and $N_\mathrm{Iter}^\mathrm{(Total)}$
are the average CG iterations for one trajectory (for generating initial pseudofermion fields, molecular dynamics,
and their sum respectively).
}
\label{table:test1}
\end{center}
\end{table}
\section{Concluding remarks}\label{sec:conclusion}
In this work, we presented one-flavor method for HMC.
The largest difference of the method presented here from RHMC and PHMC is that 
the pseudofermion action yields the one-flavor determinant without any approximations.
For the overlap fermion, the difference from the method in Ref\,\cite{DeGrand:2006ws} is that this can be used even if GWR is not exact.
For the lattice fermions with $\gamma_5$ symmetry,
it is always possible to construct real (hermitian) pseudofermion action,
but one has to careful about the positive-definiteness,
since it depends on the type of lattice fermions used.
For chirally symmetric fermions like domain-wall/overlap fermions, positive-definiteness is assured in the entire mass parameter region which can be used in two-flavor simulations.
On the other hand, for Wilson fermions, 
there exist a bound for the smallest mass where the positive-definiteness is assured. 
If the bound is larger than the mass value which is intended to use, one has to add Hasenbusch mass preconditioner then one can push the mass smaller.
Comparison between two-flavor and $(1 + 1)$-flavor using domain-wall type fermion HMC simulation shows
that
one can increase the step size of $(1 + 1)$-flavor simulation while keeping same acceptance ratio. 
The reason might be that 
for $(1 + 1)$-simulation it is effectively the same as using four-dimensional operator 
and the force from the bulk mode is completely cancelled between the light fermion and Pauli-Villars field, 
while for the two-flavor pseudofermion action used in the comparison in this work, the cancellation was done only partly.
The bottle neck of this method is generating pseudofermion field from Gaussian noise and it should be tuned and improved. 
Not only HMC itself, one can use these pseudofermion action for reweighting method,
for example, to adjust the strange quark mass or to see the effects due to the difference of up down sea quark mass by using existing configurations.
This approach for one-flavor simulation should be investigated further numerically and theoretically.

\section*{Acknowledgment}
This work is supported by 
National Center for Theoretical Sciences
(Nos. 099-2811-M-009-029, 99-2112-M-009-004-MY3).
We thank the hospitality of National Taiwan University, at which this work was started.
We thank T-W. Chiu, B. Smigielski and C.-J. D. Lin for comments and discussions.

\appendix
\section{The Eigenvalue Relation}
We prove a relation between "eigenvalues of the real part" and "the real part of the eigenvalues" for a general matrix $M$.
\beq
\lambda_{\min} (\re[M]) \le \mathrm{Re} [ \lambda( M ) ] \le \lambda_{\max} ( \re[ M ] ), \label{eq:ineq} 
\eeq
where $ \lambda( M ) $ can be any one of the eigenvalues of $ M $,
$\re [ M ]$ is the real part of $M$, $\re [ M ] \stackrel{\rm def}{=}(M + M^\dagger)/2$, 
$ \lambda_{\min} (\re[M]) $ and 
$ \lambda_{\min} (\re[M]) $ are the smallest and the largest eigenvalue of $\re[ M ]$, respectively. 
\subsubsection*{Proof}
The eigenvectors $ \phi_i $ of $ \re[ M ] $ form an orthonormal basis,
\beq
\re[ M ] \phi_i = h_i \phi_i,
~~( \phi_i , \phi_j ) = \delta_{i,j}.
\eeq
Here, $ (\,,\,) $ means the inner product.

The eigenvectors $ \psi_i $ of $ M $ can be expressed in terms of the $\phi_i$'s as,
\beq
M \psi_i = \lambda_i \psi_i,
~~
\psi_i = \sum_l c^{(i)}_l \phi_l.
\label{eq:psi}
\eeq
Here, we set $ ( \psi_i , \psi_i ) = 1 $. 
This yields $ \sum_l | c_l^{(i)}|^2 = 1 $.
The inner product of $\psi_i$ and $ M \psi_i$ gives the eigenvalue $\lambda_i$,
\beq
\lambda_i = ( \psi_i, M \psi_i ).
\label{eq:pmp}
\eeq
Then, we divide $M$ into the real part $ \re [ M ] $ and the imaginary part $\im [ M ] \stackrel{\rm def}{=} (M-M^\dagger)/2$,
and substitute these into the Eq.\,(\ref{eq:pmp}),
\beq
\lambda_i
=
( \psi_i, \re[M] \, \psi_i ) + 
( \psi_i, \im[M] \, \psi_i ). 
\eeq
After taking the real part of this equation, only the first term remains,
\beq
\mathrm{Re} ( \lambda_i ) = 
( \psi_i, \re[ M ] \, \psi_i ).
\label{eq:re_lambda}
\eeq
By substituting Eq.\,(\ref{eq:psi}) into the right hand side of this equation, one obtains
\beq
( \psi_i, \re [ M ] \, \psi_i ) 
= \sum_{k,l} \left ( c_k^{(i)} \right )^{*} c_l^{(i)} ( \phi_k, \re [ M ] \, \phi_l ) 
=  \sum_{l} \left | c_l^{(i)} \right |^2 h_l.
\eeq
This is equal to or larger than smallest eigenvalue of $ \re [ M ] $,
\beq
\sum_{l} \left | c_l^{(i)} \right |^2 h_l - \lambda_{\min}( \re [ M ] ) = 
\sum_{l} \left | c_l^{(i)} \right |^2 \lp h_l - \lambda_{\min}( \re [ M ] ) \rp \ge 0 .
\label{eq:h_subt}
\eeq
Thus, 
by using Eq.\,(\ref{eq:re_lambda}) to Eq.\,(\ref{eq:h_subt}),  
it is proven that $ \mathrm{Re} [ \lambda( M ) ] $ cannot be smaller than $\lambda_\mathrm{min}( \re [ M ] )$,
\beq
\lambda_{\min}( \re [ M ] ) \le \re [ \lambda( M ) ].
\eeq
The other inequality in Eq.\,(\ref{eq:ineq}), $ \re [ \lambda( M ) ] < \lambda_\mathrm{max}( \re [ M ] ) $, can be proven in a similar way.

\section{ Eigenvalues and Eigenvectors of Matrix $ \Delta_\pm(m) $}
Here, we derive the eigenvectors and the eigenvalues of $\Delta_{+}(m)$ defined in Eq.\,(\ref{eq:delta_m}).
We work explicitly on $N_s = 4$. After obtaining the answers for $N_s = 4$, to convert the answers for general $N_s$ is straightforward. 

Using the matrix $M_+(m,c)$, i.e.,
\beq
M_+(m,c) = ( {\bf 1} - L_+ ) ( {\bf 1} + c L_+ )^{-1} =
\frac{ 1 + \frac{1}{c}}{ {\bf 1} + c L_+ } - \frac{{\bf1}}{c},
\eeq
the matrix $\Delta_+(m_1, m_2) \stackrel{\rm def}{=} R_5 [ M_+(m_2) - M_-(m_1) ] $ is written as,
\bea
\Delta_+(m_1,m_2) & = &
R_5 
\frac{1 + \frac{1}{c}}{1 + c^4 m_2}
\left (
\begin{array}{rrrr}
   1 &  c^3 m_2 & -c^2 m_2 &  c   m_2 \\
-c   &  1       &  c^3 m_2 & -c^2 m_2 \\
 c^2 & -c       &  1       &  c^3 m_2 \\
-c^3 &  c^2     & -c       &  1
\end{array}
\right )
- 
R_5 
\frac{1 + \frac{1}{c}}{1 + c^4 m_1}
\left (
\begin{array}{rrrr}
   1 &  c^3 m_1 & -c^2 m_1 &  c   m_1 \\
-c   &  1       &  c^3 m_1 & -c^2 m_1 \\
 c^2 & -c       &  1       &  c^3 m_1 \\
-c^3 &  c^2     & -c       &  1
\end{array}
\right )
\nn
& = & g( m_1 , m_2 ,  c ) Q(c) \nonumber
\eea
Here, $g(m_1,m_2,c)$ and $Q(c)$ are defined as,
\beq
 g(m_1 , m_2 ,c) = \frac{(1 + c)(m_2 - m_1) }{( 1 + c^4 m_1 )( 1 + c^4 m_2 )},
~~~ 
Q = 
\begin{pmatrix}
 ~~c^6 & - c^5 & ~~c^4 & - c^3 \\
 - c^5 & ~~c^4 & - c^3 & ~~c^2 \\
 ~~c^4 & - c^3 & ~~c^2 & - c \\
 - c^3 & ~~c^2 & - c & ~~ 1
\end{pmatrix}
\label{eq:def_g}
\eeq
The matrix $\Delta_+(m)$ in Eq\,(\ref{eq:delta_m}) is given by $\Delta_+(m) = \Delta_+( m , 1 )$.

Eigenvalues and eigenvectors of $Q$ are, 
\beq
Q \, u_i = 0~~( i = 1, 2, 3),~~ Q \, v = \lambda_c v~~\rm{with}~~\lambda_c = c^6 + c^4 + c^2 + 1, 
\label{eq:eigen_q}
\eeq
\beq
u_1 = 
\begin{pmatrix}
1 \\
c \\
0 \\
0   
\end{pmatrix}
,~
u_2 = 
\begin{pmatrix}
0 \\
1 \\
c \\
0   
\end{pmatrix}
,~
u_3 = 
\begin{pmatrix}
0 \\
0 \\
1 \\
c   
\end{pmatrix}
,~
v = 
\frac{1}{\sqrt{ c^6 + c^4 + c^2 + 1 }}
\begin{pmatrix}
- c^3 \\
~~c^2 \\
- c \\
~~1   
\end{pmatrix}.~
\label{eq:def_v}
\eeq
Define unitary matrix $ S $ as
\beq
S = 
\begin{pmatrix}
 u_1'~
 u_2'~
 u_3'~ 
 v
\end{pmatrix}.
\label{eq:def_shat}
\eeq
Here, $u_1'$, $u_2'$ and $u_3'$ are the vectors made by orthonormalization of $u_1$, $u_2$ and $u_3$.
Then, $\Delta_+(m_1,m_2)$ is written as,
\beq
\Delta_+(m_1,m_2) = g(m_1,m_2,c) \,\, S \,\, \mathrm{diag}( 0, 0, 0, \lambda_c ) \,\, S^\dagger
 = g'(m_1,m_2,c) \,\, S \,\, \mathrm{diag}( 0, 0, 0, 1 ) \,\, S^\dagger, 
\label{eq:rwt_delta}
\eeq
with $g'(m_1,m_2,c) = \lambda_c \, g(m_1,m_2,c) $.

The similar relation for $ \Delta_-(m_1,m_2) = R_5 [ M_-(m_2) - M_-(m_1) ] $ is given by multiplying $R_5$ from left and right to this equation. 

\section{Exactness - Point of View with Eigenvalues}
Consider the eigenvalues and eigenvectors of the operator $ W_H $.
\beq
W_H \psi_i = h_i \psi_i
\eeq
The Gaussian noise field $ \Xi $, and pseudo fermion field $ \phi $,
are expressed as linear combination of the eigenvectors $ \psi_i $. 
\beq
\Xi = \sum_i b_i \psi_i, ~~
\phi = \sum_i c_i \psi_i
\eeq
The path integral of $ \phi $ is written as the product of integral w.r.t 
the coefficient $ c_i $.
\beq
\int d \phi^\dagger d \phi e^{ - \phi^\dagger \frac{1}{W_H} \phi }
=
\int \prod_i d c_i^* d c_i e^{ - \sum_i c_i^* \frac{1}{h_i} c_i }.
\eeq
When $ b_i $ has a distribution
$ e^{ - b_i^* b_i } $,
one can make the distribution proportional to
$
e^{ - c_i^* \frac{1}{h_i} c_i }
$
by multiplying by $\sqrt{h_i}$, 
$
c_i = \sqrt{h_i} b_i
$.

On the other hand, constructing  
$
c_i = f_\mathrm{app}(h_i) b_i,
$
the distribution become
$
e^{ - c_i^* \frac{1}{f_\mathrm{app}(h_i)^2} c_i }
$

The difference can be adjusted by  
accept/reject step with the probability
\beq
e^{ - c_i^* \left ( \frac{1}{h_i} - \frac{1}{f_\mathrm{app}(h_i)^2} \right )  c_i }.
\eeq
To take this factor as a probability, this must be less than one,
i.e., $f_\mathrm{app}(h_i) \ge \sqrt{h_i} $.
This condition is understood that  
$
e^{ - c_i^* \frac{1}{f_\mathrm{app}(h_i)^2} c_i }
$
has a slightly broader distribution than 
$
e^{ - c_i^* \frac{1}{\sqrt{h_i}} c_i }
$,
and the accept/reject step makes the distribution narrow by suppressing larger $ c_i $.

\section{Linear Algebra}
Here, we remind the readers some relations of linear algebra. 
\begin{itemize}
\item
Schur Decomposition
\beq
\begin{pmatrix}
A & B \\
C & D
\end{pmatrix}
=
\begin{pmatrix}
1       & 0 \\
CA^{-1} & 1
\end{pmatrix}
\begin{pmatrix}
A & 0 \\
0 & D - C A^{-1} B
\end{pmatrix}
\begin{pmatrix}
1 & A^{-1} B \\
0 & 1
\end{pmatrix}
=
\begin{pmatrix}
1 & B D^{-1} \\
0 & 1
\end{pmatrix}
\begin{pmatrix}
A - B D^{-1} C & 0 \\
0 & D
\end{pmatrix}
\begin{pmatrix}
1        & 0 \\
D^{-1} C & 1
\end{pmatrix}
\eeq

\item
The Inversion of an Schur Complement
\beq
\begin{pmatrix}
0 \\ y 
\end{pmatrix}
=
\begin{pmatrix}
A & B \\
C & D
\end{pmatrix}
\begin{pmatrix}
w \\ x 
\end{pmatrix}
~\Rightarrow~
y = ( D - CA^{-1} B ) \,x
\eeq
\item
The Determinant of an Operator with Projector $P_{\pm}$
\beq
\det P'_+ M P'^\dagger_+ = \det (P_+ M + P_-)
= \det (M P_+ + P_-) 
\eeq
\beq
\det (M P_+ + P_-) 
= \det( M \,( P_+ + M^{-1} P_-))
= \det( M ) \det ( P_+ + M^{-1} P_-)
= \det(M) \det (P'_- M^{-1} P'^\dagger_-)
\eeq
\end{itemize}



\bibliographystyle{model1-num-names}



\end{document}